
%
\input harvmac
\noblackbox
%
%

\def\apm{\alpha^{\prime}}

\def\IP{\relax{\rm I\kern-.18em P}}
%
%
\lref\klem{A. Klemm and R. Schimmrigk, Nucl. Phys. {\bf B411} (1994) 559.}
\lref\pert{I. Antoniadis, S. Ferrara, E. Gava, K. S. Narain and
T. R. Taylor hep-th/9504034, B. DeWit, V. Kaplunovsky, J. Louis
and D. Lust, hep-th/9504006.}
\lref\lerche{A. Klemm, W. Lerche, S Yankielowicz and S. Theisen,
Phys. Lett. {\bf B344} (1995) 169. }
\lref\asyk{A. Strominger, Phys. Rev. Lett. {\bf 55}, (1985) 2547. }
\lref\cfg{S. Cecotti, S. Ferrara and L. Girardello,Int. J. Mod. Phys. {\bf A4}
(1989) 2475.}
\lref\bbs{K. Becker, M. Becker and A. Strominger, unpublished.}
\lref\dlwp{B. de Wit, P. Lauwers and A Van Proeyen,
Nucl. Phys. {\bf B255} (1985) 269.}
\lref\wit{E. Witten,
hep-th/9503124.}
\lref\rey{S. J. Rey, Phys. Rev. {\bf D43} (1991) 526.}
\lref\drev{For a review see M. J. Duff,  R. R. Khuri, and J. X. Lu, ``String
Solitons,'' hep-th/9412184.}
\lref\senone{A. Sen, ``Macroscopic Charged Heterotic String'', hep-th/9206016,
Nucl. Phys. {\bf B388} (1992) 457.}
\lref\sentwo{A. Sen,
hep-th/9504027.}
\lref\nati{N. Seiberg,
Nucl. Phys. {\bf B303}, (1986), 288;  P. Aspinwall and D.
Morrison,
preprint DUK-TH-94-68, IASSNS-HEP-94/23,
hep-th/9404151.}
\lref\duffss{M. J. Duff,
NI-94-033,
CTP-TAMU-49/94, hep-th/9501030.}
\lref\kv{S. Kachru and C. Vafa,
hep-th/9505105.}
\lref\sen{A. Sen, Int.J.Mod.Phys. {\bf A9}  (1994) 3707,  hep-th/9402002.}
\lref\sentform{A. Sen,
Phys. Lett. {\bf B329} (1994) 217, hep-th/9402032.}
\lref\swit{N. Seiberg and E. Witten,
hep-th/9407087, Nucl. Phys. {\bf B426}, (1994), 19.}
\lref\vafa{C. Vafa, unpublished.}
\lref\ferr{A. Ceresole, R. D'Auria and S. Ferrara,
Phys. Lett. {\bf B399}, (1994), 71;
A. Ceresole, R. D'Auria, S. Ferrara and A. Van Proeyen,
hep-th/9502072.}
\lref\schsen{J. Schwarz and A. Sen, Phys. Lett. {\bf B312} (1993) 105,
hepth/9305185; P. Binetruy, Phys. Lett. {\bf B315} (1993) 80, hep-th/9305069. }
\lref\DAHA{ A. Dabholkar and J. A. Harvey,
Phys. Rev. Lett.
{\bf 63} (1989) 478.}
\lref\DGHR{ A. Dabholkar, G. Gibbons, J. A. Harvey and F. R. Ruiz,
``Superstrings and Solitons, '' Nucl. Phys. {\bf B340} (1990) 33.}
\lref\hetsol{ A. Strominger,
Nucl. Phys. {\bf B343} (1990) 167; E: Nucl. Phys. {\bf B353} (1991) 565.}
\lref\sswt{ A. Strominger,  ``Superstrings with Torsion '',
Nucl. Phys. {\bf B274} (1986) 253.}
\lref\towne{P.K.  Townsend,
R/95/2,
hep-th/9501068. }
\lref\bcov{M. Bershadsky, S. Cecotti, H. Ooguri and C. Vafa, Comm. Math. Phys.
{\bf 165} (1994) 311; Nucl. Phys. B{\bf405} (1993) 279. }
\lref\vcon{C. Vafa, hep-th/9505023.}
\lref\town{C.M. Hull and P.K.  Townsend,
QMW-94-30, R/94/33,
hep-th/9410167. }
\lref\snone{N. Seiberg, hep-th/9402044.}
\lref\bhole{G. Horowitz and A. Strominger,
Nucl. Phys. {\bf B360} (1991) 197.}
\lref\wbran{C. G. Callan, J. A. Harvey and A. Strominger,
Nucl. Phys. {\bf B367} (1991) 60.}
\lref\tei{R. Nepomechie, Phys. Rev. {\bf D31} (1985) 1921;
C. Teitelboim, Phys. Lett. {\bf B176} (1986) 69.}
\lref\wsheet{C.G. Callan, J. A. Harvey and A. Strominger,
``Worldsheet Approach to Heterotic Instantons and  Solitons,'' Nucl. Phys. {\bf
359}
(1991) 611. }
\lref\SCH{J. H. Schwarz, Nucl. Phys. {\bf B226} (1983) 269. }
\lref\coneheads{A. Strominger, hep-th/9504090; B. Greene, D. Morrison and A.
Strominger, hep-th/95040145.}
\lref\hs{J. A. Harvey and A. Strominger,
hep-th/9504047.}
\lref\vwt{C. Vafa and E. Witten, hep-th/9505053.}
\lref\SWII{N. Seiberg and E. Witten,
Nucl. Phys. {\bf B431} (1994) 484;  hep-th/9408099.}
\lref\mirrbook{{\it Essays on Mirror Manifolds,} Ed. S.-T. Yau, (International
Press,
Hong Kong 1992). }
\lref\barth{For details on $K3$ and the Enriques surface see W. Barth, C.
Peters and
A. Van de Ven, {\it Compact Complex Surfaces}, (Springer-Verlag, Berlin,
1984).}
\lref\distgr{J. Distler and B. R. Greene, Nucl. Phys. {\bf B309} (1988) 295.}

%
%

\Title{\vbox{\baselineskip12pt
\hbox{EFI-95-26, HUTP-95/A019}\hbox{CERN-TH 95/131}\hbox{UCLA/95/TEP/17}
\hbox{hep-th/9505162}}}
{\vbox{\centerline{\bf{SECOND-QUANTIZED MIRROR SYMMETRY} }}}
{
\baselineskip=12pt
 \centerline{Sergio Ferrara\foot{CERN, 1211 Geneva 23, Switzerland;
Department of Physics,University
of California,Los Angeles,CA 90024-1547},~
Jeffrey A. Harvey\foot{Enrico Fermi Institute, 5640 Ellis Avenue, University of
Chicago,
Chicago, IL 60637 USA},~
Andrew Strominger\foot{Department of Physics, University of California,
Santa Barbara, CA 93106-9530 USA}~ and ~
Cumrun Vafa\foot{Lyman Laboratory of Physics, Harvard University,
Cambridge, MA 02138 USA}}

\bigskip \bigskip
\centerline{\bf Abstract}
We propose and give strong evidence for
a duality relating Type II theories on Calabi-Yau
spaces and heterotic strings on $K3 \times T^2$, both of which have
$N=2$ spacetime
supersymmetry.  Entries in the dictionary relating the
dual theories are derived from an analysis of the soliton string worldsheet
in the context of $N=2$ orbifolds of dual $N=4$
compactifications of Type II and heterotic strings. In particular we
construct a pairing between Type II string
theory on a self-mirror Calabi-Yau space $X$ with
$h^{11}= h^{21}= 11$ and a $(4,0)$
background of heterotic string theory on $K3 \times T^2$.
Under the duality transformation the usual first-quantized mirror symmetry
of $X$  becomes a second-quantized mirror symmetry which
determines nonperturbative quantum effects. This enables us to
show that the quantum moduli space for this example agrees with
the classical one. Mirror symmetry of $X$ implies
that the low-energy $N=2$ gauge theory is finite, even at
enhanced symmetry points. This prediction
is verified by direct computation on the
heterotic side.  Other
branches of the moduli
space, which are not finite $N=2$ theories,
are connected to this one via black hole condensation.  }
\Date{May, 1995}
%
\eject
\newsec{Introduction}
 It has been conjectured that many string theories in various dimensions
have dual descriptions\refs{\DAHA,\hetsol,\town,\duffss, \towne, \wit, \vafa}
which
allow one to obtain information about the theory at strong coupling by
performing
weak coupling calculations in the dual theory. Currently the most
compelling evidence exists for
a string-string duality relating the IIA theory on $K3$ to the heterotic
string on $T^4$\refs{\nati,\town,\wit,\sentwo,\hs,\vwt},
which is a compactified version of string-fivebrane duality \hetsol .
When these theories are further reduced to four dimensions  by toroidal
compactification,  six-dimensional string-string duality implies $S$-duality
of the resulting four-dimensional theories\refs{\schsen, \duffss, \wit}.
These four-dimensional theories
have $N=4$ supersymmetry. $S$-duality in these theories is the natural
generalization to supergravity and superstring theory of $S$-duality
in $N=4$ Yang-Mills theory.

It has also been realized recently that duality plays a fundamental role
in understanding the dynamics of gauge theories with $N=2$ \swit\ and $N=1$
\snone\ supersymmetry.  It is natural to wonder whether duality in these
theories
might also have roots in duality in string theory.  Since $N=2$ gauge
theories are under the most precise control, this seems like a natural starting
point for investigating this idea.

String theories in four dimension with $N=2$ supersymmetry arise either
by compactification of Type II theories on Calabi-Yau spaces or by
compactification of the heterotic string on $K3 \times T^2$.  The question
then is whether it is possible to find a duality relating Type II theory
on a particular Calabi-Yau space to a heterotic string background on
$K3 \times T^2$. In trying to make such a correspondence, a puzzle immediately
arises.
For $N=2$ compactification of a type IIA string, $\chi=2(N_v-N_h+1)$,
where $\chi$ is the Euler character of the Calabi-Yau space $X$ and
$N_h= h^{21}(X)+1$ ($N_v= h^{11}(X)$)
is the number of four dimensional hyper (vector) multiplets.
On the other hand, it is easy to see that $2(N_v-N_h+1)$ is not constant
over all branches of the moduli space of $K3 \times T^2$ heterotic string
compactifications:
it can change at enhanced symmetry points\foot{Of course this is just the
classical picture. The connectivity of the various
branches of the moduli space
could differ quantum mechanically.}. Since $\chi$ is
constant over the moduli space of a single Calabi-Yau space, this cannot be
represented by a type II string on a single Calabi-Yau space. Rather one
requires
a family of Calabi-Yau spaces. The type II description of changing
$2(N_v-N_h+1)$ by
passing through an enhanced symmetry point will  involve jumping from one
Calabi-Yau space to its neighbor via black hole condensation, as recently
described in \coneheads.

Given such a dual pair of string theories, the exact quantum moduli
space can be determined from classical computations. The reason for this
is that the dilaton is in a vector (hyper) multiplet on the
heterotic (type IIA) side. Supersymmetry prevents couplings between
neutral vector and
hypermultiplets  in the low-energy effective
action \dlwp. Therefore the vector (hyper) multiplet geometry on the
type IIA (heterotic) side can not depend on the string coupling, and is
exact at tree level. The full quantum geometry is thus determined by
a classical vector (hyper) multiplet computation in the type IIA (heterotic)
representation of the theory. In particular spacetime instanton
effects in the
heterotic representation are worldsheet
instantons in the dual type IIA representation, and so
may be classically computed.
This is second-quantized mirror symmetry.
Second-quantized mirror symmetry can be used to sum up
spacetime instanton corrections just as
ordinary first-quantized mirror symmetry can be used to sum up
worldsheet instanton corrections \mirrbook.

In this paper we will investigate in detail a specific
branch of the $N=2$ moduli space.  Although we will find
a precise mapping and correspondence between dual theories
in this example the general situation is not yet understood.
We begin with the dual $N=4$ pair
consisting of a IIA theory on $K3 \times T^2$ and the heterotic theory
on $T^6$. The IIA theory has a soliton string\refs{\sentwo,\hs} (a
compactification of
the fivebrane of \wbran ) whose effective worldsheet theory is exactly that
of a heterotic string on $T^4$\hs. Next we take a $Z_2$ orbifold of the IIA
theory,
yielding an $N=2$ Calabi-Yau compactification on a manifold $X$. The
corresponding
orbifold on the heterotic side is then determined from the $Z_2$
action on the soliton string. The dual pair so constructed
indeed have the same values of $N_h$ and $N_v$.

The fact that $X$ is its
own mirror is used to argue that there are no worldsheet or spacetime instanton
corrections on this branch of the moduli space. This implies that the
effective $N=2$ gauge theories at all enhanced symmetry points must lie
in finite $N=2$ representations, a prediction which is verified by
direct computation. For example we find the finite $SU(2)$, $N_f=4$ theory, and
our construction gives new insights into the beautiful results of \SWII.
For example, the triality discovered in \SWII\ arises as the dual image of
a $T$-duality transformation in the type II theory.

The suggestion of a duality relation between $N=2$ heterotic and type II
compactifications was made previously in \refs{\ferr, \lerche}. Perturbative
aspects of $N=2$ heterotic compactifications were studied in
\pert. Other examples of this duality including stringy
analogs of Seiberg-Witten monopole points have been found recently in \kv.

\newsec{The Calabi-Yau Construction}

We wish to construct a Calabi-Yau space  as an orbifold of $K3 \times T^2$.
As discussed in the introduction, we will obtain a specific test of
string duality by constructing a Calabi-Yau space which
is self-mirror.  We will show that this is the case for
a Calabi-Yau space $X$ constructed as a freely acting orbifold of $K3 \times
T^2$. Type II string propagation on
$X$ has
only $N=2$ spacetime supersymmetry rather than $N=4 $ spacetime supersymmetry.
Such an orbifold can be constructed by
utilizing a well-known
freely acting involution $\theta_1$ of certain $K3$ surfaces.
For $K3$ surfaces admitting this involution the quotient is known as an
Enriques surface \barth. Let $z_3$ be a complex coordinate on $T^2$. We now mod
out by
$\Theta = \theta_1 \theta_2$ where $\theta_2$ acts as inversion on $T^2$,
$\theta_2 z_3 \theta_2^{-1} = -z_3$.  The Hodge numbers of the Enriques surface
are $h_{11}=10, ~~h_{02}=h_{20}=0$.
In particular there is no holomorphic two-form. So the holomorphic two-form
$\Omega$ on $K3$ is odd under $\theta_1$ and thus the holomorphic three-form
$\Omega \wedge dz_3$ is preserved by $\Theta$ and the quotient is a
Calabi-Yau space with Euler number zero. It is important to note that the
involution
exists and this construction can
be carried out on a large subspace of the $K3$ moduli space. Generic points in
this
subspace do not contain any degenerations or enhanced symmetries.

Now by the Lefschetz fixed point theorem we have
\eqn\one{ \chi_{\theta_1} =0 = 2+ Tr_2 \theta_1 }
where $\chi_{\theta_1}$ is the Euler character of the fixed point set of
$\theta_1$, the $2$ comes from $H^0(K3)$ and $H^4(K3)$ and
$Tr_2 \theta_1$ is the trace
of the action of $\theta_1$ on $H^2(K3)$.  Thus $\theta_1$ has
eigenvalues $[(-1)^{12}, (+1)^{10} ]$ acting on $H^2(K3)$. Furthermore,
since $\theta_1$ is $-1$ on $H^{2,0}(K3)$ and $H^{0,2}(K3)$ we learn that
on $H^{1,1}(K3)$ $\theta_1$ has eigenvalues $[(-1)^{10}, (+1)^{10}]$.
Now we can compute $h^{11}$ of this Calabi-Yau space. We get
$10$ from the $10$ $(1,1)$ forms on $K3$ with eigenvalue $1$ and
we get one more from the $(1,1)$ form on $T^2$ which is clearly even
under the involution. Thus $h^{1,1}= 11$.  Similarly for $h^{21}$ we get
$10$ from taking the wedge product of $10$ $(1,1)$ forms with eigenvalue
$-1$ on $K3$
with the $(1,0)$ form on $T^2$ which is odd under $\theta_2$ and
we get one more from taking the wedge product of the $(2,0)$ form
on $K3$ and the $(0,1)$ form on $T^2$. Thus $h^{21}=11$ which of
course was required by $\chi=0$. The low-energy theory has gauge
group $U(1)^{12}$ with one of the $U(1)$ factors being the
graviphoton, and $N_v=11$ and $N_h=12$ (including the dilaton).

The action of the involution $\theta_1$ on $K3$ can be usefully summarized
as follows. The intersection form on $K3$ is
\eqn\kper{L_{IJ}=\left[\Gamma_8 \oplus \Gamma_8 \oplus \sigma^1\oplus
\sigma^1\oplus \sigma^1\right]_{IJ} , \qquad \sigma^1 = \pmatrix{
0 & 1 \cr
1 & 0 \cr
}}
where $\Gamma_8$ is the Cartan matrix for $E_8$.   $\theta_1$ acts by
interchanging two copies of $\Gamma_8 \oplus \sigma^1$ and as
$-1$ on the third $\sigma^1$. This will be useful in the following section
when we construct the corresponding orbifold on the heterotic side.

Note that we can choose a point in the moduli space of $K3$ corresponding to
$T^4/Z_2$
and perform the same $Z_2$ modding out of $(T^4 / Z_2) \times T^2$
described above.  In such
a description, the $Z_2$ acts by reflection on $T^2$ and its action
on $T^4$ is $(-1,1)$ on the complex coordinates
$(z_1,z_2)$ of $T^4$ as well as a shift by a half lattice
vector in both $z_1$ and $z_2$ directions.  Then the untwisted
sector contributes $(3,3)$ to $(h^{11},h^{21})$ and one
of the twisted sectors (corresponding to the first $Z_2$)
contributes $(8,8)$.

An important modification of this construction is as follows.
The fundamental group of this Calabi-Yau space is $Z_2$. This means we can
give a Wilson line expectation value to the RR U(1) field which is present
in the ten-dimensional IIA theory. This has no effect whatsoever on
string perturbation theory, since all fundamental string states are
neutral under this U(1). However we will argue later that
the theory without the Wilson line is nonperturbatively inconsistent
due to problems with ``black hole level-matching''. We define $X$
to be the Calabi-Yau space including the RR Wilson line.

The string tree level moduli space of $X$ can be computed locally\foot{We
ignore the
issue of global identifications in most of this paper.}. The moduli
space of complex structures is given by the special geometry
\eqn\sv{{\cal M}_V={SU(1,1)\over U(1)} \times {SO(10,2)\over SO(10)\times
SO(2)}.}
The first (second) factor is the space of complex structures on $T^2$
(Enriques surface). In the IIA theory these hypermultiplet moduli are augmented
by
RR scalars which also fill out hypermultiplets. The enlarged hypermultiplet
moduli space is
determined at tree level from \sv\ and the c-map \cfg\ as the
quaternionic space
\eqn\sh{{\cal M}_H={SO(12,4)\over SO(12)\times SO(4)} .}
The moduli space of complexified Kahler forms on $X$ is easily
computed to leading-order in sigma-model perturbation theory
from the cubic intersection form on $H^2(X)$. The intersection of
the two form on $T^2$ with the $\theta_1$ even cohomology of
$K3$ is $2(\Gamma_8\oplus \sigma^1)$. It then follows from the
formulae in \asyk\ that the moduli space of vector multiplets is
the special geometry ${\cal M}_V$ given in \sv\ above.

Since $K3$ and $T^2$ are both their own mirrors, and we are modding
out by a free $Z_2$ action, it
is natural to expect that  $X$ is also
its own mirror.   This is basically true\foot{
More precisely $X$ is mirror to itself with a $Z_2$
discrete torsion turned on, which does not affect the local
geometry of moduli space, similar to the situation studied in \ref\amori{P.
Aspinwall and D. Morrison, Phys. Lett. {\bf 334B} (1994) 79.}.}
as follows
from the analysis of similar orbifolds in
\ref\vwdi{C. Vafa and E. Witten, Jour. Geom. and Phys. {\bf 15}
(1995) 189.}.
This provides a method to compute the exact Kahler moduli
space using the inverse c-map from ${\cal M}_H$ and the
fact that ${\cal M}_H$ does not receive world-sheet
instanton corrections \distgr. This reproduces the
leading-order sigma-model result quoted above. We conclude that
there can be no worldsheet instanton corrections to the Kahler moduli
space, and the exact tree level moduli space is given by
${\cal M}_V \times {\cal M}_H$.

\newsec{Soliton Strings and the Duality Dictionary}
We now need to know how to map orbifold constructions on the IIA side
into orbifold constructions on the heterotic side.  String-string
duality tells us that the symmetries of the $K3$ moduli
given by $SO(20,4;Z)$ are transformed to the same group for
the heterotic theory, which implies that we should
think of the heterotic lattice as the integral cohomology
lattice of $K3$ and that the Poincare duality on $K3$ cohomology
is what gives the left-right decomposition of the Narain lattice
($*=1$ subspace is right-movers and $*=-1$ is the left-movers).
Any discrete symmetry of  a $K3$ manifold will be a symmetry of its cohomology
lattice and will thus, through this
identification, give us the action on the heterotic side, up
to potential phases which are represented by `shift vectors' on the heterotic
side.
This identification of the heterotic lattice with the  integral $K3$ cohomology
lattice, as required by string duality,
has been found to hold in a physically beautiful way \refs{\hs,\sentwo}\
realizing the heterotic string as a soliton in the
IIA theory.

The soliton string is a fivebrane \wbran\ with four of its spatial extensions
wrapping $K3$.
The worldsheet fields of the soliton string arise as zero modes
of the soliton solution. We will not repeat the entire discussion of
\hs, but rather mention the relevant points. In the notation of
\hs\ the zero modes are
\eqn\ansatz{ C =  {\apm \over 2 \pi} X^I(\sigma) U_I(y) \wedge
d e^{2\phi_0-2 \phi(x)} ,}
with
$e^{2 \phi(x)}$ the
background dilaton field of the string soliton and $I= 1,...22$.
The harmonic two-forms
$U_I$ comprise
an integral basis for $H^2(K3,Z)$ with intersection form given by
\kper.
The $X^I$s are worldsheet fields subject to the chiral constraint
\eqn\rwr{ \partial_\pm X^I=\pm{H^I}_J\partial_\pm X^J,}
where $H$ relates the $U$ basis to its Hodge dual
\eqn\hdl{ {\hat \ast} U_I=U_J{H^J}_I.}
$H$ has signature $(19,3)$ (and depends on the modulus of the $K3$
surface) so there are 19 left-moving and 3 right-moving $X^I$s.
An additional zero mode involves
a combination of $C$ and the
one-form potential $A$
\eqn\azero{\eqalign{A&=  {1 \over 2 \pi} X^0(\sigma) de^{2 \phi_0-2\phi},\cr
C&=-{1 \over \pi} X^0(\sigma) e^{2 \phi_0 -2\phi} H,\cr}}
This  gives
an additional $(1,1)$ (left,right)-moving bosonic zero mode. The gauge
transformation law
is $\delta A=d\epsilon,~~
\delta C=-2\epsilon H$, so at zero worldsheet momentum \azero\ corresponds to a
gauge
transformation with asymptotic parameter $X^0 \over 2 \pi$. Quantization of RR
charge
then implies periodicity of  $X^0$, which we take to be $X^0 \sim X^0 +2 \pi$.
Finally
there
is one complex bosonic zero mode $z^3(\sigma)$ corresponding to transverse
motion in the $T^2$
factor and two more real zero modes (which will not enter in to the discussion)
from transverse motion in four-dimensional Minkowski space. Together with the
right-moving superpartners this comprises the worldsheet content of a
$T^6$ compactified heterotic string.

Now we translate the action of $\theta_1$ into an action on the
zero mode coordinates of the heterotic string.
As discussed below \kper, the action of $\theta_1$ exchanges two
sets of 10 $U_I$'s with intersection matrix $\Gamma_8\oplus \sigma^1$
and reverses the sign of two more with intersection matrix $\sigma^1$ .
But this
is equivalent, from \ansatz, to exchanging two
sets of 10 $X^I$'s with momenta living in a lattice $\Gamma_8\oplus \sigma^1$
and reversing the sign of two more $X^I$'s living on the lattice $\sigma^1$.
In string theory it is conventional to enlarge the intersection form by working
with the full even-dimensional cohomology, that is by including $H^0$
and $H^4$ as well. This enlarges the intersection form \kper\  by an
additional $\sigma^1$ factor.
If a RR Wilson line is included,  $\Theta$ is accompanied by a  RR U(1) gauge
transformation whose square is the identity. The zero momentum part of $X^0$ is
a
gauge transformation so inclusion of the Wilson line corresponds to the $Z_2$
shift
$X^0\rightarrow X^0+\pi$ which can be thought of as a shift in the additional
$\sigma^1$ arising from $H^0$ and $H^4$.

\newsec{The $N=2$ Heterotic String Orbifold}

We now wish to describe this as a conventional orbifold of the
heterotic string.
Working at a general point in the Narain moduli space consistent with
the action discussed above we  start with an even, self-dual
Lorentzian lattice $\Gamma^{(22,6)}$  of
the form
\eqn\biglat{\Gamma^{(9,1)} \oplus \Gamma^{(9,1)} \oplus \Gamma^{(1,1)} \oplus
\Gamma^{(1,1)} \oplus
\Gamma^{(2,2)} }
where the two $\Gamma^{(9,1)}$ factors are isomorphic.
We then  mod out by a $Z_2$ action which exchanges the first two factors in
\biglat,
acts as $-1$ on the third and fifth factors and as a shift in the fourth
factor
if we include a RR Wilson line in the type II theory.
The $-1$ action on the fifth factor of \biglat\ is
inversion of the coordinate $z^3$ on $T^2$.
Thus the heterotic orbifold is a $Z_2$ orbifold of a special
Narain compactification which has twelve negative eigenvalues on the left and
four on the right.

Now with no shifts the vacuum energy for this twist is $-1/4$ on the left and
zero on the right and thus does not satisfy level-matching. This can be
rectified by including a $Z_2$ shift on the invariant $\Gamma^{(1,1)}$ of the
form
$\delta = (p_L,p_R)/2$ where $p^2 = p_L^2 - p_R^2 = 2$.  This shifts the
left-moving vacuum energy by $p_L^2/8$ and the right-moving vacuum energy
by $p_R^2/8$. Since the difference in vacuum energies is now zero
level-matching
and hence
modular invariance are satisfied. We presume that on the IIA side this arises
as a non-perturbative consistency condition once an analog of modular
invariance is understood for black hole states.

At generic points  in the lattice \biglat\ the massless spectrum of the
orbifold
including the shift consists of
eleven vector multiplets and twelve hypermultiplets, as predicted by duality.
To see this we work in the Ramond-Neveu-Schwarz formalism.
Then the right-moving vacuum in the bosonic
sector consists of four states with the quantum numbers of  the bosonic states
of a vector supermultiplet with eigenvalue $+1$ under the twist and four states
with the quantum numbers of the bosonic states of a hypermultiplet with
eigenvalue $-1$ under the twist. Combining these states with the twelve
left-moving
massless states with eigenvalues $-1$ under the twist yields the twelve
hypermultiplets.
The other states yields the eleven  vector multiplets and the gravitational
multiplet.
The analogous counting goes through for the fermionic states. At generic
points there are no massless states in the twisted sector because the
right-moving
vacuum energy is shifted up by $p_R^2/8$ which is non-zero at generic points.
At tree-level the moduli space is given locally by ${\cal M}_V \times
{\cal M}_H$, as follows from the general structure of Narain moduli.

\newsec {The Exact Moduli Space and Finite N=2 Theories}
 We have found that at tree level the moduli space is given by
${\cal M}_V \times
{\cal M}_H$ in  both the heterotic and type II representations.
On the heterotic side there are no quantum corrections to
hypermultiplets, while on the type II side there are no quantum corrections
to vector multiplets. Therefore by duality the moduli space is exact in
all expansion parameters.

The absence of worldsheet instanton corrections on the type II side
implies the absence of spacetime instanton corrections on the heterotic
side\foot{Other dual pairs of type II heterotic theories constructed
by going through enhanced symmetry points will in general have
non-trivial instanton corrections. Analysis of these corrections will
provide a further stringent  consistency check on our proposal.} .
This is possible only if all the Yang-Mills beta functions vanish
at all enhanced symmetry points. Thus it must be the case that the
states come down to zero mass only in finite representations\foot{The
gravitational beta functions do not in general vanish. These control
higher dimensional $R^2$
corrections to the leading low-energy effective action of the type
studied in \refs{\bcov,\vcon}. Analysis of these effects will lead to yet
further checks of our proposed duality.}. Let us
check this.

There are two types of enhanced symmetry points which arise
at special points in \biglat. The first arises by going to an enhanced
symmetry point in $\Gamma^{(9,1)}$ which in general yields a rank
nine simply-laced non-abelian gauge symmetry (e.g $E_8 \times SU(2)$)\foot{
For example at the moduli of $X$ corresponding to the orbifold point of $K3$
mentioned earlier we
get an $SU(2)^8$ non-abelian
gauge symmetry.}.
Now at such a point it is clear that the symmetric combination of lattice
vectors has eigenvalue $+1$ under the interchange and yields a
vector supermultiplet in the adjoint representation.  Note that
the resulting world-sheet current algebra is now at level two.
The antisymmetric
combination of lattice vectors also transforms in the adjoint representation
of the group but since it is odd under the $Z_2$ it gives a hypermultiplet.
This of course is the field content of a finite $N=2$ gauge theory.

The other type of enhanced symmetry point yields a more intricate finite
theory. We can also go to an enhanced symmetry point in the $\Gamma^{(1,1)}$
lattice containing the shift $\delta$ by going to the self-dual radius. At this
point lattice vectors are of the form $(p_L,p_R)= (m+n,m-n)/\sqrt{2}$ with
$m,n$ integers and the shift vector is $\delta = (1/\sqrt{2},0)$ which is half
of a root vector of $SU(2)$ .  In the untwisted sector we find a vector
supermultiplet
in the adjoint representation of $SU(2)$ from lattice points $(\pm \sqrt{2},0)$
and including the oscillator structure.
In the twisted  sector we now have massless states from points
with ${\tilde p}_L^2= 1/2$ and ${\tilde p}_R^2 = 0$ with ${\tilde p}_L =
(m+n+1)/\sqrt{2}$
and ${\tilde p}_R = (m-n)/\sqrt{2}$. There are two such states with $m=n=0$ or
$m=n=-1$ and these two states transform as a doublet of $SU(2)$.  We next
need to compute the multiplicity of these states. There are two contributions
coming from the left and right-moving vacuum degeneracies. On the
right we now have four anti-periodic and four periodic fermions. Quantization
of the zero modes yields a four-fold degenerate ground state and the GSO
projection reduces this to two. In addition the $Z_2$ action on
$\Gamma^{(1,1)} \oplus \Gamma^{(2,2)}$ has eight fixed points which
gives a total degeneracy of $16$.
One can check, using the number of fixed points for asymmetric orbifolds
\ref\asym{K.S. Narain, M.H. Sarmadi and C. Vafa, Nucl. Phys. {\bf B356} (1991)
163; Nucl. Phys. {\bf B288} (1987) 551.},
 that the interchange of the $\Gamma^{(9,1)}$ factors does not
give any additional degeneracy. This  degeneracy is precisely that of
four hypermultiplets in the fundamental representation of $SU(2)$. As is
well known, this is also a finite $N=2$ theory which was previously studied in
\SWII.  Some aspects of the duality for this theory found in \SWII\ can also
be found in this construction.

One striking aspect of the duality in this context was the observation
in \SWII\
that the $SL(2,Z)$ symmetry of the gauge coupling constant in this example
acts also on the flavor symmetry which in this example is $Spin(8)$.
However, a subgroup of $SL(2,Z)$, $\Gamma(2)$,
acts only on the coupling constant and does not act on the flavors.
Moreover the extra elements of  $SL(2,Z)$ given by reduction
mod $2$ form the permutation
group $S_3$ on three objects, which acts as triality on the
$Spin(8)$ representations.  Let us see whether
we can see any hints of this structure in our model.

First we note that $\Gamma(2)$ is the subgroup of
$SL(2,Z)$ preserving the spin structures on the torus
and the four
conjugacy classes of $Spin(8)$
can be identified with the four fixed points of a $T^2$ under the
involution with $S_3$ permuting three of the fixed points
(it does not act on the one at the origin).

According to the dictionary of string-string duality
applied to compactifications on a $T^2$ down to 4 dimensions
\refs{ \wit, \duffss}\ the dilaton of the heterotic theory
gets mapped to the Kahler class of $T^2$ on the type II side.
Here we have divided further by a $Z_2$ acting as reflection
on $T^2$ and an involution on $K3$.  Since the $Z_2$ involution
on $T^2$ exists for all complex and Kahler structures, this
in particular means that the Kahler class of $T^2$ survives
the modding out, and it thus is still parametrized by
the fundamental domain of the  torus modulus.  This
means that the heterotic string coupling constant
should likewise still be parametrized by the fundamental
domain of the upper-half plane.  However, we are asking
for more information.  In particular $SL(2,Z)$ transformations
on the coupling constant should also act as triality as described
with only
the $\Gamma(2)$ subgroup  leaving the flavors untouched.
Note that, due to mirror symmetry on $X$ this is equivalent to the same
question about the $SL(2,Z)$ which acts on the complex structure of $T^2$
which is more easily realized geometrically.  Note that the 4 hypermultiplets
of $SU(2)$ on the heterotic side came from the four fixed points of $T^2$.
Given that $T^2$ is common to both type II and heterotic part, it
thus follows that the subgroup of $SL(2,Z)$ that does not act
on the flavors, i.e. does not reshuffle the fixed points, is just
$\Gamma (2)$, as expected from \SWII .  In order to
verify that the extra elements of $SL(2,Z)$ do act as triality
on the flavor representations, we have to study more carefully
how the $Spin(8)$ is realized on the fixed points of the asymmetric
orbifold of the heterotic theory.  In particular physical
vertex operators are particular
 linear combinations of fixed point sets \refs{\asym}.  A careful
study  following the above
strategy should lead to the triality action of \SWII .  As
a further check on these ideas, note that at the other
type of enhanced gauge symmetry points (e.g. $E_8\times SU(2)$)
the matter does not come from the fixed point sets and thus there
is no action of $SL(2,Z)$ on the
flavor symmetry, as expected from $N=4$
duality.

Finally, we note that the $N_f=4$, $SU(2)$ theory also has a Higgs branch
along which the $SU(2)$ doublets condense and break the $SU(2)$ gauge
symmetry.  Since we start out with 4 doublets of $SU(2)$, this
leaves us after breaking $SU(2)$ with $8-3=5$ extra hypermultiplets.
Moreover we have gotten rid of one vector multiplet (corresponding
to the $U(1)$ of $SU(2)$).
 In the Type IIA theory this should correspond to black holes becoming
massless and condensing.  Assuming this new phase corresponds to a Calabi-Yau
compactification,  it would have to be one with Hodge numbers
$(h^{11},h^{12})=(10,16)$. Such a space (in fact its mirror)
indeed occurs in the list of \klem,
where it is denoted $(7,7,8,8,12,12,18)$.
This Calabi-Yau space is not self-mirror and should have world-sheet
instanton corrections on the Type IIA side and spacetime quantum corrections
on the heterotic side which  are related by second-quantized
mirror symmetry.

\bigskip
\centerline{\bf Acknowledgements}\nobreak
We would like to thank E. Witten
for participation at the initial stages of this work.
We are grateful to K. Becker, M. Becker, S. Kachru, R. Schimmrigk
and especially D. Morrison for
useful discussions.
This work was supported in part
by NSF Grants No.~PHY 91-23780,~PHY-92-18167, DOE Grant No. DOE-91ER40618,
Grant DE-FGo3-91ER40662,TASK C and
EEC Science program SC1 CI92-0789.

\listrefs
\end